\documentclass[12pt,english,floatfix,nofootinbib,superscriptaddress,aps,prd,preprint]{revtex4}
\usepackage[utf8]{inputenc}
\usepackage{float}
\usepackage{array}
\usepackage{lipsum}
\usepackage{graphicx}
\usepackage{amsmath,amsthm,amsfonts,amssymb}
\usepackage{graphicx}
\usepackage[english]{babel}
\usepackage{xcolor}
\usepackage{tensor}
\usepackage{esint}
\usepackage[dvips]{epsfig}
\usepackage[dvips]{graphicx}
\usepackage{float}
\usepackage{units}
\usepackage{textcomp}
\usepackage{mathrsfs}
\usepackage{amsmath}
\usepackage[makeroom]{cancel}
\usepackage{amssymb}
\usepackage{amsbsy}
\usepackage{amsfonts}
\usepackage{amssymb,mathrsfs,xcolor}
\usepackage{esint}
\usepackage{array}
\usepackage{graphicx}
\usepackage{mathtools,slashed}
\usepackage{slashed}
\usepackage{hyperref}
\hypersetup{
    colorlinks,
    citecolor=blue,
    filecolor=green,
    linkcolor=purple,
    urlcolor=red,
}

\newcommand{\ie}{\begin{equation}}
\newcommand{\fe}{\end{equation}}
\newcommand{\se}{\begin{eqnarray}}
\newcommand{\ff}{\end{eqnarray}}

\newcommand{\nslash}{\slashed{\nabla}}
\newcommand{\Aslash}{\slashed{A}}

\newcommand{\calO}{\mathcal{O}}
\newcommand{\up}{\uparrow}
\newcommand{\down}{\downarrow}

\usepackage{hyperref}
\hypersetup{colorlinks,breaklinks,
			citecolor=[rgb]{0,0.0,1.0},
            urlcolor=[rgb]{0.0,0.0,1.0},
            linkcolor=[rgb]{0,0.5,0.9}}

\begin{document}

\title{Nonlocal Dirac theory in curved spacetime and the Bunch--Davies vacuum}

\author{P. J. Porf\'{\i}rio}
\email{pporfirio@fisica.ufpb.br}
\affiliation{Departamento de Física, Universidade Federal da Paraíba, Caixa Postal 5008, 58051-970, João Pessoa, Paraíba,  Brazil} 

\author{Ramires N. da Silva}
\email{rns2@academico.ufpb.br }
\affiliation{Departamento de Física, Universidade Federal da Paraíba, Caixa Postal 5008, 58051-970, João Pessoa, Paraíba,  Brazil}


\date{\today}

\begin{abstract}
We formulate a nonlocal spin-$1/2$ field theory in curved spacetime in which the form factor is a nonpolynomial entire function of the covariant Dirac operator. We derive the corresponding nonlocal Dirac equation and analyze its coupling to gravity and to a $\rm U(1)$ gauge field. As an application, we study the theory in the open chart of de Sitter spacetime and construct the associated nonlocal Bunch--Davies modes. The nonlocal dynamics can be mapped, branch by branch, onto the local Dirac equation with an effective mass $M_n$. For the exponential form factors considered here, the effective masses are expressed in terms of branches of the Lambert-$W$ function. We show that the Euclidean analyticity structure of the Bunch--Davies construction is preserved. On the physical positive-residue branch, the Bogoliubov coefficients, squeezing parameter, occupation number, and entanglement entropy acquire explicit nonlocal corrections, while the absence of physical fermionic supercurvature modes is preserved. The local theory is smoothly recovered in the limit $\Lambda\to\infty$.
\end{abstract}

\maketitle

\section{INTRODUCTION}
Nonlocality has long been explored as a possible extension of the standard quantum field theory (QFT) framework, particularly when a model's ultraviolet (UV) behavior becomes problematic. In such theories, nonlocality means that the action involves operators with infinitely many derivatives, typically encoded through analytic form factors acting on the fundamental fields. Early studies of nonlocal interactions date back to \cite{Wataghin,Pais,Pauli,Hayashi}, while a systematic quantum formulation was developed by Efimov \cite{Efimov}, who showed that suitable nonlocal structures can yield UV-finite amplitudes in scalar field theories. Related ideas were subsequently explored in Abelian and non-Abelian gauge theories~\cite{Krasnikov,Moffat}, nonlocal realizations of the Higgs mechanism~\cite{Modesto4,Hashi}, scalar electrodynamics, where quantum corrections were studied explicitly \cite{Gama}, and supersymmetric theories formulated in superspace~\cite{Gama:2017ets,Gama:2026ude}.

Motivations for nonlocality also arise beyond conventional QFT. In string theory, an infinite tower of higher-derivative contributions appears in the low-energy effective description through $\alpha'$ corrections, with $\alpha'$ denoting the inverse string tension \cite{Kaku,Green,Witten:1985}. Nonlocal structures are likewise intrinsic to $p$-adic string models \cite{Freund,Freund2,Moeller}. These examples suggest that infinite-derivative operators may encode effective remnants of a more fundamental microscopic theory rather than merely phenomenological modifications.

The gravitational sector provides another important setting in which nonlocality has been extensively studied. Since general relativity is perturbatively nonrenormalizable, higher-curvature operators naturally arise once quantum corrections are included \cite{Shapiro}. Although such higher-derivative terms improve the UV behavior, they introduce additional propagator poles, some associated with negative-norm ghost states \cite{Hawking, Antoniadis}. Infinite-derivative gravity was therefore proposed as a means to improve the UV behavior while avoiding extra pathological degrees of freedom \cite{Tomboulis,Modesto1,Modesto2,Modesto3,Modesto5}. This framework has been applied to nonsingular and nonlocal cosmologies \cite{Biswas:2005qr,Biswas,Jhingan,Biswas2,Dimitrijevic,Modesto8,Koshelev,Koshelev2}, modified black-hole geometries \cite{Modesto6,Modesto7}, and other exact configurations \cite{Buoninfante1,Buoninfante2,Andrade,Nascimento,Ghoshal:2022mnj}.

A key ingredient in many ghost-free nonlocal constructions is the use of nonpolynomial entire functions of the kinetic operator. Thus, form factors of this type introduce no zeros of their own and can, under suitable conditions, improve the UV behavior without modifying the propagator pole structure. Related constructions have been investigated in scalar theories~\cite{Briscese,Eugenio, Cleverson, Cesar:2026ljk}, gauge and supersymmetric theories~\cite{Gama2,Gama3}, and nonlocal gravity~\cite{Boos,Biswas2011}.

In contrast to the scalar, gauge, and gravitational sectors, comparatively little attention has been devoted to genuinely nonlocal theories of spin-$1/2$ fields. Existing fermionic constructions, such as those in \cite{Krasnikov,Moffat}, often introduce nonlocality through form factors depending on the d'Alembertian. Although admissible, this choice does not fully exploit the first-order nature of fermionic dynamics, which is governed by the Dirac operator $\slashed{\partial}=\gamma^\mu\partial_\mu$ rather than directly by $\Box$.

This point was emphasized in \cite{Nascimento2026-ie}, where nonlocality was introduced directly through the Dirac operator by replacing the local kinetic term with $\slashed{\partial}f(\slashed{\partial})$. Since the form factor then depends on a first-order, matrix-valued differential operator, the analytic structure of the fermionic propagator becomes particularly relevant, including its pole structure, the possible emergence of additional branches, and the physical interpretation of the corresponding degrees of freedom.

In the present work, we extend this nonlocal spin-$1/2$ framework to curved spacetime by replacing the ordinary Dirac operator with its generally covariant counterpart and defining the nonlocal form factor as an entire function of this operator. This provides a covariant framework for investigating nonlocal fermionic dynamics in gravitational and cosmological backgrounds.

The paper is organized as follows. In Sec.~\ref{sec2}, we formulate the free nonlocal spin-$1/2$ theory in curved spacetime and discuss
its coupling to a $\rm U(1)$ gauge field. In Sec.~\ref{sec3}, we specialize to the open chart of de Sitter spacetime and construct the corresponding nonlocal Bunch--Davies (BD) modes. In Sec.~\ref{sec4}, we analyze the nonlocal BD vacuum and observables, identify the physical principal branch of Model II, and determine the corresponding corrections to particle occupation and entanglement entropy. Finally, Sec.~\ref{sec5} summarizes our conclusions.

\section{Nonlocal spin-$1/2$ field theory in curved spacetime}\label{sec2}

\subsection{Free theory}

To extend the nonlocal spin-$1/2$ model of \cite{Nascimento2026-ie} from Minkowski to curved spacetime, we begin with the local Dirac action minimally coupled to gravity,
\begin{equation}\label{Dirac}
    S=\int d^{4}x\,\sqrt{-g}\left[\frac{i}{2}e^{\mu}_{\,\,a}\left(\bar{\Psi}\gamma^{a} \nabla_{\mu}\Psi-(\nabla_{\mu}\bar{\Psi})\gamma^{a}\Psi\right)-m\bar{\Psi}\Psi\right],
\end{equation}
where $e^{a}_{\,\,\mu}(x)$ is the vierbein field. The metric can be locally defined in terms of a set of anholonomic bases: frames, $\left\{e_{a}(x) \right\}$, and co-frames, $\left\{\theta^{a}(x) \right\}$,
\begin{equation}
g=\eta_{ab}\theta^{a}\otimes \theta^{b}=g_{\mu\nu}dx^{\mu}\otimes dx^{\nu}.
\end{equation} 

The duality condition between the frame and co-frame gives $g_{\mu\nu}(x)=e_{\mu}^{\,\,a}(x)e_{\nu}^{\,\,b}(x)\eta_{ab}$. The vierbein and its inverse relate the flat- and curved-space Dirac matrices through: $\gamma_{\mu}=e_{\mu}^{\,\,a}\gamma_a$ and $\gamma^{\mu}=e^{\mu}_{\,\,a}\gamma^{a}$.  The  covariant derivatives acting on spinors are defined by
\begin{eqnarray}
\nabla_{\mu}\Psi=\partial_{\mu}\Psi+\Gamma_{\mu}\Psi,\qquad \nabla_{\mu}\bar{\Psi}=\partial_{\mu}\bar{\Psi}-\bar{\Psi}\Gamma_{\mu},
\end{eqnarray}
with the spinorial connection defined by $ \Gamma_{\mu}=-\dfrac{i}{4}\omega_{\mu ab}\sigma^{ab}$, where $\omega_{\mu ab}$ are the components of the spin connection and $\sigma^{ab}=\dfrac{i}{2}[\gamma^{a},\gamma^{b}]$ are the generators of the covering Lorentz group in the spinor representation.

Up to a boundary term, Eq.~(\ref{Dirac}) can be written in the usual first-order form. Following \cite{Nascimento2026-ie}, we introduce nonlocality through an entire function of the covariant Dirac operator,
\begin{equation}\label{eq1c}   
S=\int d^4x\sqrt{-g}\left[i\bar\Psi\slashed{\nabla}f(\slashed\nabla) \Psi-m\bar\Psi\Psi\right],
\end{equation}
where $\slashed{\nabla}\equiv \gamma^{\mu}\nabla_\mu$ is the Dirac operator in a curved spacetime.

Varying the action in Eq.~\eqref{eq1c} with respect to $\bar{\Psi}$ yields the nonlocal Dirac equation. Multiplying it from the left by $i\slashed{\nabla}f(\slashed{\nabla})+m$, we obtain its squared form,
\begin{equation}\label{nonl}
    [i\slashed{\nabla}f(\slashed\nabla)-m]\Psi=0 \qquad \Longrightarrow  \qquad [\slashed{\nabla}^2f^2(\slashed\nabla)+m^2]\Psi=0.
\end{equation}

We adopt the metric signature $(+ - - -)$, the Clifford algebra $\{\gamma^\mu,\gamma^\nu\}=2g^{\mu\nu}\hat{\rm{I}}$ and $R^{\rho}{}_{\sigma\mu\nu}= \partial_{\mu}\Gamma^{\rho}_{\nu\sigma} - \partial_{\nu}\Gamma^{\rho}_{\mu\sigma} + \Gamma^{\rho}_{\mu\lambda}\Gamma^{\lambda}_{\nu\sigma} - \Gamma^{\rho}_{\nu\lambda}\Gamma^{\lambda}_{\mu\sigma}$. With these conventions, the Lichnerowicz identity reads
\begin{equation}\label{b2}
 \begin{aligned}
        \slashed{\nabla}^{2} &=\frac{1}{2}\gamma^{\mu}\gamma^{\nu}\left([\nabla_{\mu},\nabla_{\nu}]+\{\nabla_{\mu},\nabla_\nu\}\right) =\Box-\frac{R}{4},
\end{aligned}
\end{equation}
where we have used the following relations \cite{Toms, BirrellDavies, ParkerToms}: $[\nabla_{\mu},\nabla_{\nu}]\Psi=\dfrac{1}{4}R_{\mu\nu ab}\gamma^{a}\gamma^{b}\,\Psi$ and $R_{\mu\nu ab}\gamma^{\mu}\gamma^{\nu}\gamma^{a}\gamma^{b}=-2R$. Using this result, the nonlocal Dirac equation~\eqref{nonl} can be rewritten as 
\begin{equation} \label{eqrr}
    \left[\mathcal{K}f^2(\slashed\nabla)+m^2\right]\Psi=0,
\end{equation}
where $\mathcal{K} \equiv \Box- R/4$. Let us consider the spectrum of eigenvalues of the squared Dirac operator $\slashed{\nabla}^{2}$, namely,
\begin{eqnarray}\label{eigenvalue}
    \slashed{\nabla}^2 \Psi=\mathcal{K}\Psi=-\lambda^2 \Psi,
    \label{eigen}
\end{eqnarray}
where $-\lambda^2$ is the eigenvalue of $\slashed{\nabla}^2$. Repeated action of $\slashed{\nabla}$ on Eq.~\eqref{eigen} yields the recursive relations
\begin{eqnarray}\label{2m}
\slashed{\nabla}^{2j+1}\Psi=(-1)^j\lambda^{2j}\slashed{\nabla}\Psi,\qquad \slashed{\nabla}^{2k}\Psi=(-1)^k\lambda^{2k}\Psi,\qquad j, k \in \mathbb{N}_0.
\end{eqnarray}

Moreover, if $f(\slashed{\nabla})$ is entire and zero-free in the complex plane, the Weierstrass factorization theorem allows it to be written as $f(\slashed{\nabla})=e^{g(\slashed{\nabla})}$, with $g(\slashed{\nabla})$ entire~\cite{AhlforsComplex,Complex}. Hence,
although the form factor introduces no zeros of its own, additional propagator poles may arise from the zeros of the full fermionic kinetic operator in the presence of the mass term, giving rise to branch-dependent solutions. Since $f^2(\slashed{\nabla})$ is also
entire, it admits a power-series expansion. Separating its even and odd powers and using Eq.~(\ref{2m}), we obtain
\begin{equation}
f^2(\slashed{\nabla})\Psi = \left[ \hat{\rm{I}}\sum_{k=0}^{\infty}(-1)^k c_{2k} \frac{\lambda^{2k}}{\Lambda^{2k}} + \slashed{\nabla}\sum_{k=0}^{\infty}(-1)^k c_{2k+1} \frac{\lambda^{2k}}{\Lambda^{2k+1}} \right]\! \Psi,
\label{f22}
\end{equation}
where $\Lambda$ denotes the characteristic nonlocality scale, assumed to lie in the high-energy regime. Equation~(\ref{f22}) shows that, on an eigenspace of $\slashed{\nabla}^{\,2}$, the form factor reduces to the two-dimensional operator basis $[\hat{\rm{I}},\slashed{\nabla}]$. The local limit $\Lambda\to\infty$ requires $f(0)=1$, and hence $c_0=1$ in the expansion of $f^2$. This result agrees with the corresponding Minkowski-space expression obtained in \cite{Nascimento2026-ie}.

The squared nonlocal Dirac equation, Eq.~\eqref{eqrr}, has a scalar-like structure reminiscent of nonlocal Klein--Gordon equations \cite{FrolovNonlocalKG,IzadiMoayedi}. However, the form factor depends on the first-order Dirac operator, as explicitly shown in Eq.~\eqref{f22}, whereas the d'Alembertian enters through $K=\Box-R/4$. Hence, the spinor components do not generally satisfy independent scalar nonlocal equations.

The Hermiticity of the nonlocal kinetic operator imposes an additional constraint on the admissible form factors, which was not considered in \cite{Nascimento2026-ie}. With the standard Dirac inner product and suitable boundary conditions, the covariant Dirac operator is formally anti-Hermitian, $\slashed{\nabla}^{\ddagger}=-\slashed{\nabla}$, so that $i\slashed{\nabla}$ is formally self-adjoint. For an analytic form factor $f(\slashed{\nabla})$, the kinetic operator $i\slashed{\nabla}f(\slashed{\nabla})$ is therefore formally self-adjoint provided that
\begin{equation}\label{6}
f(\slashed{\nabla})^{\ddagger}=f(\slashed{\nabla}),
\end{equation}
which corresponds to the reality condition $f(z)^*=f(-z^*)$. This requirement is independent of the condition that $f$ be entire and zero-free. Thus, although a zero-free entire form factor introduces no zeros of its own, a standard Hermitian Lorentzian formulation further requires the above reality condition.

\subsubsection{Exponential form factors}
We consider two exponential entire form factors that generalize those introduced in \cite{Nascimento2026-ie}:
\begin{eqnarray}
 \text{Model I}&:   \label{f1}f_{1}(\slashed{\nabla})=e^{-\slashed{\nabla}_{\!\!\Lambda}}\,,\\
 \text{Model II}&:   \label{f2}f_{2}(\slashed{\nabla})=e^{-i\slashed{\nabla}_{\!\!\Lambda}}\,,
\end{eqnarray}
where $\slashed{\nabla}_{\Lambda}\equiv \slashed{\nabla}\!/\!\Lambda$. Expanding $f_{1}$ as a power series and using Eq.~\eqref{2m}, we obtain
\begin{eqnarray}
       f_{1}(\slashed{\nabla})= \sum \limits_{j=0}^{\infty} \frac{(-1)^j \,\lambda^{2j}}{(2j)!\Lambda^{2j}} \hat{\rm{I}}-\frac{1}{\lambda} \sum \limits_{k=0}^{\infty} \frac{(-1)^k \,\lambda^{2k+1}}{(2k+1)!\Lambda^{2k+1}}\slashed{\nabla}= \cos{\left(\lambda/\Lambda\right)}\hat{\rm{I}}-\frac{1}{\lambda}\sin{\left(\lambda/\Lambda\right)}\slashed{\nabla},
    \label{fa3}
\end{eqnarray}
consequently
\begin{equation}\label{f1quadrado}
   f_{1}^{\,2}(\slashed{\nabla})= \cos{\left(2\lambda/\Lambda\right)} \hat{\rm{I}}-\frac{1}{\lambda}\sin{\left(2\lambda/\Lambda\right)}\slashed{\nabla}.
\end{equation}

Similarly, for Model II, 
\begin{equation}\label{f2quadrado}
     f_{2}^{\,2}(\slashed{\nabla})= \cosh{\left(2\lambda/\Lambda\right)}\hat{\rm{I}}-\frac{i}{\lambda}\sinh{\left(2\lambda/\Lambda\right)}\slashed{\nabla},
\end{equation}
which can then be inserted into the nonlocal Dirac equation, Eq.~\eqref{eqrr}. Notice that the trigonometric and hyperbolic structures in Eqs.~(\ref{f1quadrado}) and (\ref{f2quadrado}) are interchanged relative to \cite{Nascimento2026-ie}. This interchange follows from the spectral convention given by Eq.~(\ref{eigen}), associated here with our choice of metric and curvature conventions.

\subsection{Coupling to a $\rm U(1)$ gauge field}
We minimally couple a $\rm U(1)$ gauge field to the nonlocal spin-$1/2$ theory~\eqref{eq1c} through the replacement $\slashed{\nabla} \to \slashed{\mathcal{D}} =\slashed{\nabla}+i\rm{e} \slashed{A}$, where $\rm{e}$ denotes the fermionic charge. The interacting action then takes the form
\begin{equation}
S_{I}=\int d^4x\sqrt{-g} \bar\Psi\left[i\slashed{\mathcal{D}}\,f(\slashed{\mathcal{D}})-m \right]\Psi.
\label{eq2c}
\end{equation}

Under the local gauge transformations $\Psi'=\rm U\Psi$, $\bar{\Psi}'=\bar{\Psi}\rm U^{-1}$, and $A'_\mu=A_\mu+\partial_\mu\alpha$, with
$\rm U=e^{-i\rm{e}\alpha}$, the gauge-covariant Dirac operator transforms as $\slashed{\mathcal D}'=\rm U\slashed{\mathcal D}\rm U^{-1}$. Since $f$ is analytic, $f(\slashed{\mathcal D}')=\rm U \mathit{f}(\slashed{\mathcal D})\rm U^{-1}$, and hence
\begin{equation}
\bar{\Psi}'\slashed{\mathcal D}' f(\slashed{\mathcal D}')\Psi' = \bar{\Psi}\slashed{\mathcal D} f(\slashed{\mathcal D})\Psi, \qquad \bar{\Psi}'\Psi'=\bar{\Psi}\Psi.
\end{equation}

Therefore, the nonlocal action in Eq.~\eqref{eq2c} is gauge invariant (see also \cite{Nascimento2026-ie}) and yields the interacting nonlocal Dirac equation, whose squared form is
\begin{equation}
        [\slashed{\mathcal{D}}^{2} f^{2}(\slashed{\mathcal{D}})+m^{2}]\Psi=0,
    \label{kleinb}
\end{equation}
where
\begin{equation}
\begin{aligned}
\slashed{\mathcal{D}}^{2} &=\left(\mathcal{D}^{2}-R/4\right)\hat{\rm{I}}+\frac{\rm{e}}{2} \sigma^{\mu\nu}\rm F_{\mu\nu}\\
&=[\mathcal{K} + i\rm{e} (\nabla_{\mu}A^{\mu} + 2A_{\mu}\nabla^{\mu}) - \rm{e}^{2}A_{\mu}A^{\mu}]\,\hat{\rm{I}}+\frac{\rm{e}}{2} \sigma^{\mu\nu}F_{\mu\nu},   
\end{aligned}
    \label{b2}
\end{equation}
and $\rm F_{\mu\nu}=\partial_{\mu}A_{\nu}-\partial_{\nu}A_{\mu}$ is the field strength.

Unlike in the free theory, $\slashed{\mathcal D}^2$ contains the spin--field coupling $\dfrac{\rm{e}}{2}\sigma^{\mu\nu}F_{\mu\nu}$ and is therefore not generally proportional to the identity in spinor space. Consequently, the free spectral decomposition in Eqs.~\eqref{2m} cannot be applied directly. We therefore treat the gauge field perturbatively, in analogy with the flat-space analysis of \cite{Nascimento2026-ie}.

For definiteness, we illustrate the perturbative interacting construction for Model II, Eq.~(\ref{f2}); the corresponding expansion for Model I follows analogously. To express the interacting nonlocal Dirac equation (\ref{kleinb}) in terms of the spectrum of $\slashed{\nabla}$, we apply the Zassenhaus expansion \cite{Zassenhaus,CasasZassenhaus}
\begin{equation}\label{formfactor}
    f_2^{\,2}(\slashed{\mathcal{D}})= e^{-2i \slashed{\mathcal{D}}_{\!\Lambda}}=e^{2\rm{e}\Aslash_{\!\Lambda}} \,e^{-2i\nslash_{\!\!\Lambda}} \, e^{2i\rm{e}[\Aslash_{\!\Lambda},\nslash_{\!\!\Lambda}]} \cdots\,,
\end{equation}
where
\begin{equation}
 [\Aslash_{\!\Lambda},\nslash_{\!\!\Lambda}] =\frac{1}{\Lambda^2}\left(-2i\sigma^{\mu\nu}A_{\mu}\nabla_{\nu}-  \nabla^{\mu}A_{\mu} + i\frac{\sigma^{\mu\nu}}{2}\rm F_{\mu\nu}\right).
\end{equation}

Since the interacting nonlocal Dirac equation is treated perturbatively, we expand only the explicitly nonlocal contributions in the large-$\Lambda$ regime. More precisely, we restrict the analysis to spectral modes and gauge-field configurations for which $
\dfrac{|\lambda|}{\Lambda}$, $\dfrac{\rm{e}|A|}{\Lambda}$, and $\dfrac{\rm{e}|F|}{\Lambda^{2}}\ll1$, where $\lambda$ is the spectral parameter defined in Eq.~(\ref{eigenvalue}). These conditions are understood for the field configurations considered, in a fixed gauge. The gauge-dependent exponential and the commutator factors generated by the Zassenhaus expansion can then be consistently expanded in inverse powers of $\Lambda$, while the free nonlocal factor $f_{2}^{\,2}(\slashed{\nabla})=e^{-2i\nslash/\Lambda}$ is kept in its spectral form. At the same time, the local gauge-covariant structure contained in $\slashed{\mathcal D}^{\,2}$, including terms such as $\rm{e}^2A_\mu A^\mu$, is retained without performing a separate truncation in powers of the gauge field,
\begin{eqnarray}\label{pertubative}
\Big[\! \big(\mathcal{D}^{2}-R/4+ \frac{\rm{e}}{2}\sigma^{\mu\nu}\rm F_{\mu\nu}\big)\big(
        1+2\rm{e}\Lambda^{\!-1}\Aslash + \calO (\! \Lambda^{\!-2} )\big) \mathit{f}_{2}^{\,2}(\nslash) \big( 1+\calO (\!\Lambda^{\!-2})\big)\! + \textit{m}^2\Big]\Psi=0.
\end{eqnarray}

The Zassenhaus expansion generates nonstandard interactions involving the spinor, gauge field, curvature, and their derivatives through nested commutators such as $[\nslash,\Aslash]$, $[\nslash,[\nslash,\Aslash]]$, $[\Aslash,[\nslash,\Aslash]]$ $\cdots$ . Using Eq.~\eqref{b2}, separating the conventional gauge-covariant contributions from the explicitly nonlocal corrections, the interacting equation (\ref{pertubative}) can be written, at the order considered, in the schematic form
\begin{align}
\Big[\mathcal{K} f_{2}^2(\nslash) +m^2 + \left( i\rm{e}\left(\nabla_\mu A^\mu + 2A^\mu\nabla_\mu\right) -\rm{e}^2 A_\mu A^\mu + \frac{\rm{e}}{2}\sigma^{\mu\nu}\rm F_{\mu\nu} \right) f_{2}^{\,2}(\nslash)  +\cdots \Big]\Psi = 0,
\end{align}
where the ellipsis denotes the explicitly nonlocal corrections. For Model II, these begin at $\mathcal O({\rm e}/\Lambda)$ through the gauge-dependent factor
$2{\rm e}\Aslash/\Lambda$, while the first commutator contribution appears at $\mathcal O({\rm e}/\Lambda^2)$. Higher-order terms contain the nested commutators generated by the Zassenhaus expansion and couple the spinor field to the gauge field, curvature, and their derivatives.

\section{NONLOCAL DIRAC THEORY IN OPEN DE SITTER SPACETIME}\label{sec3}

\subsection{de Sitter space in the open chart}
We now apply the nonlocal Dirac equation to de Sitter spacetime in the open chart and construct the corresponding BD modes. The open slicing can be obtained by analytic-continuation from the Euclidean four-sphere $S^4$~\cite{Maldacena:2012xp}. Let us consider the Euclidean continuation of $dS_4$, namely the four-sphere $S^4$, embedded in five-dimensional Euclidean space,
\begin{equation}
    (X^0)^2+(X^1)^2+(X^2)^2+(X^3)^2+(X^4)^2=H^{-2},
\end{equation}
where $H^{-1}$ is the de Sitter radius and $X^{i}$, with $i=0,1,2,3,4$, are the embedding coordinates. The Lorentzian de Sitter spacetime can be embedded in five-dimensional Minkowski spacetime with line element given by
\begin{equation}
    ds_{5}^2=(dX^0)^2-(dX^1)^2-(dX^2)^2-(dX^3)^2-(dX^4)^2.
    \label{eqds}
\end{equation}

The open chart provides a particular slicing of de Sitter spacetime by negatively curved spatial hypersurfaces, corresponding to three-dimensional hyperboloids $H^3$. The open chart is obtained by choosing
\begin{align}
X^0 &= H^{-1} \sinh(Ht) \cosh r, \label{openX0}\\ X^i &= H^{-1} \sinh(Ht) \sinh r \, \omega^i, \qquad i=1,2,3, \label{openXi}\\ X^4 &= \pm H^{-1} \cosh(Ht), \label{openX4} 
\end{align}
where $\sum (\omega^i)^2 = 1$. Substituting the parametrizations in Eqs. \eqref{openX0}--\eqref{openX4} into Eq.~\eqref{eqds}, one finds the line element for $dS_4$, in open coordinates,
\begin{equation}
ds^2 = dt^2 - H^{-2}\sinh^2(Ht)dH_3^2,
\end{equation}
where $dH^2_3=dr^2 + \sinh^2 r \, d\Omega_2^2$ is the line element of the unit three-hyperboloid.

The $(\pm)$ sign in Eq.~\eqref{openX4} defines two causally disconnected open regions of $dS_4$. The plus sign corresponds to the $X^4>0$ region, denoted by $R$, whereas the minus sign corresponds to the $X^4<0$ region, denoted by $L$.

\subsection{Bunch--Davies states in the open chart}

The Euclidean-vacuum construction in open de Sitter spacetime is particularly relevant to open-inflation scenarios \cite{sasaki1995}. For the local Dirac field, the corresponding open chart construction of the Bunch--Davies vacuum was developed in \cite{Kanno2017-zv}. It proceeds by solving the Dirac equation separately in the $R$ and $L$ charts, analytically continuing the mode functions through the Euclidean section, and assembling the positive- and negative-frequency BD modes with support on both charts.

Our purpose is to extend the corresponding construction to the nonlocal Dirac equation~\eqref{nonl}. Let $\Psi_M$ denote a local Dirac mode satisfying $i\slashed{\nabla}\Psi_M=M\Psi_M$. Since the form factor is an analytic function of the Dirac operator, it acts diagonally on such a mode, i.e., $ f(\slashed{\nabla})\Psi_M=f(-iM)\Psi_M$. The nonlocal equation is therefore satisfied provided that
\begin{equation}\label{mass_condition}
M f(-iM)=m .
\end{equation}

BD modes can consequently be obtained from the corresponding local solutions by the replacement $m\to M_n$, where $M_n$ denotes a solution of Eq.~\eqref{mass_condition}.

\subsection{Open chart ingredients}

Introducing the dimensionless time coordinate $\tau=Ht$, and subsequently denoting $\tau$ again by $t$, the metric becomes
\begin{equation}
\label{metric_open}
ds^2=H^{-2}\left[dt^2-\sinh^2 t\, dH_3^2\right].
\end{equation}
The two-component spinor harmonics on $H^3$ satisfy
\begin{equation}
\label{H3_spinors}
\tilde{\nabla}\!\!\!/\;\chi^{(-)}_{p\ell m}=-ip\,\chi^{(-)}_{p\ell m}, \qquad \tilde{\nabla}\!\!\!/\;\chi^{(+)}_{p\ell m}=+ip\,\chi^{(+)}_{p\ell m},
\qquad p>0.
\end{equation}

For real branches, the corresponding open chart solutions are obtained from the local expressions by the replacement $m\to M_n$. A standard BD quantization is assigned only to real, positive-residue branches. The radial functions defined in \cite{Kanno2017-zv} then become
\begin{align}
\phi_p^{(n)}(z)&=(z^2-1)^{-3/4}\left(\frac{z+1}{z-1}\right)^{\!\!ip/2}\!\! {}_2F_1\!\left(-i\mu_{n},\,i\mu_{n},\,\frac12-ip,\,1-z^2\right), \label{phi_def}\\[2mm] \varphi_p^{(n)}(z)&=-\frac{i\mu_{n}}{2}\,\frac{(z^2-1)^{-1/4}}{\frac12-ip} \left(\frac{z+1}{z-1}\right)^{\!\!ip/2} \!\!{}_2F_1\!\left(1-i\mu_{n},\,1+i\mu_{n},\,\frac32-ip,\,1-z^2\right),
\label{varphi_def}
\end{align}
where $z\equiv\cosh t$ and $\mu_{n}\equiv M_n/H$. The negative-frequency radial functions are
\begin{align}
\phi_p^{(n)*}(z) &=(z^2-1)^{-3/4}\left(\frac{z+1}{z-1}\right)^{\!\!-ip/2} \!\!{}_2F_1\!\left(-i\mu_{n},i\mu_{n},\,\frac12+ip,\,1-z^2\right),
\label{phi_star_def}\\[2mm]
\varphi_p^{(n)*}(z) &=\frac{i\mu_{n}}{2}\,\frac{(z^2-1)^{-1/4}}{\frac12+ip} \left(\frac{z+1}{z-1}\right)^{\!\!-ip/2}\!\! {}_2F_1\!\left(1-i\mu_{n},\,1+i\mu_{n},\,\frac32+ip,\,1-z^2\right),
\label{varphi_star_def}
\end{align}
and the normalization factor is
\begin{equation}
\label{Nb_def}
N_{n}^2= 1+ e^{-2\pi p}+(1- e^{-2\pi p}) \frac{\sinh^2 (\pi \mu_n)}{\cosh^2(\pi p)} .
\end{equation}

\subsection{Explicit nonlocal Bunch--Davies modes}

For each real branch $M_n$, the eight analytically continued open chart solutions can be formally constructed by inserting the replacements above into Eqs.~(5.1)--(5.9) of Ref.~\cite{Kanno2017-zv}. These solutions admit the standard interpretation as nonlocal BD modes only when the associated propagator pole has positive residue. We list the explicit solutions below.
\subsubsection{Positive-frequency modes}
\begin{equation}
\label{mode_PR_up}
\Psi^{+(R)}_{\up,n}(z,\Omega)=
\begin{cases}
\dfrac{1}{N_n}
\begin{pmatrix}
\phi_p^{(n)}(z)\,\chi^{(-)}_{p\ell m}(\Omega)\\[1mm]
\varphi_p^{(n)}(z)\,\chi^{(-)}_{p\ell m}(\Omega)
\end{pmatrix}, & z=z_R,\\[5mm]
\dfrac{1}{N_n}
\begin{pmatrix}
\bigl(A_n\varphi_p^{(n)}(z)+B_n\phi_p^{(n)*}(z)\bigr)\chi^{(-)}_{p\ell m}(\Omega)\\[1mm]
\bigl(-A_n\phi_p^{(n)}(z)+B_n\varphi_p^{(n)*}(z)\bigr)\chi^{(-)}_{p\ell m}(\Omega)
\end{pmatrix}, & z=z_L.
\end{cases}
\end{equation}

\begin{equation}
\label{mode_PR_down}
\Psi^{+(R)}_{\down,n}(z,\Omega)=
\begin{cases}
\dfrac{1}{N_n}
\begin{pmatrix}
\varphi_p^{(n)}(z)\,\chi^{(+)}_{p\ell m}(\Omega)\\[1mm]
\phi_p^{(n)}(z)\,\chi^{(+)}_{p\ell m}(\Omega)
\end{pmatrix}, & z=z_R,\\[5mm]
\dfrac{1}{N_n}
\begin{pmatrix}
\bigl(-A_n\phi_p^{(n)}(z)+B_n\varphi_p^{(n)*}(z)\bigr)\chi^{(+)}_{p\ell m}(\Omega)\\[1mm]
\bigl(A_n\varphi_p^{(n)}(z)+B_n\phi_p^{(n)*}(z)\bigr)\chi^{(+)}_{p\ell m}(\Omega)
\end{pmatrix}, & z=z_L.
\end{cases}
\end{equation}

\begin{equation}
\label{mode_PL_up}
\Psi^{+(L)}_{\up,n}(z,\Omega)=
\begin{cases}
\dfrac{1}{N_n}
\begin{pmatrix}
\bigl(-A_n\phi_p^{(n)}(z)+B_n\varphi_p^{(n)*}(z)\bigr)\chi^{(-)}_{p\ell m}(\Omega)\\[1mm]
-\bigl(A_n\varphi_p^{(n)}(z)+B_n\phi_p^{(n)*}(z)\bigr)\chi^{(-)}_{p\ell m}(\Omega)
\end{pmatrix}, & z=z_R,\\[5mm]
\dfrac{1}{N_n}
\begin{pmatrix}
\varphi_p^{(n)}(z)\,\chi^{(-)}_{p\ell m}(\Omega)\\[1mm]
-\phi_p^{(n)}(z)\,\chi^{(-)}_{p\ell m}(\Omega)
\end{pmatrix}, & z=z_L.
\end{cases}
\end{equation}

\begin{equation}
\label{mode_PL_down}
\Psi^{+(L)}_{\down,n}(z,\Omega)=
\begin{cases}
\dfrac{1}{N_n}
\begin{pmatrix}
-\bigl(A_n\varphi_p^{(n)}(z)+B_n\phi_p^{(n)*}(z)\bigr)\chi^{(+)}_{p\ell m}(\Omega)\\[1mm]
\bigl(-A_n\phi_p^{(n)}(z)+B_n\varphi_p^{(n)*}(z)\bigr)\chi^{(+)}_{p\ell m}(\Omega)
\end{pmatrix}, & z=z_R,\\[5mm]
\dfrac{1}{N_n}
\begin{pmatrix}
-\phi_p^{(n)}(z)\,\chi^{(+)}_{p\ell m}(\Omega)\\[1mm]
\varphi_p^{(n)}(z)\,\chi^{(+)}_{p\ell m}(\Omega)
\end{pmatrix}, & z=z_L.
\end{cases}
\end{equation}

\subsubsection{Negative-frequency modes}

\begin{equation}
\label{mode_NR_up}
\Psi^{-(R)}_{\up,n}(z,\Omega)=
\begin{cases}
\dfrac{1}{N_n}
\begin{pmatrix}
\phi_p^{(n)*}(z)\,\chi^{(+)}_{p\ell m}(\Omega)\\[1mm]
-\varphi_p^{(n)*}(z)\,\chi^{(+)}_{p\ell m}(\Omega)
\end{pmatrix}, & z=z_R,\\[5mm]
\dfrac{1}{N_n}
\begin{pmatrix}
\bigl(A_n^*\varphi_p^{(n)*}(z)+B_n^*\phi_p^{(n)}(z)\bigr)\chi^{(+)}_{p\ell m}(\Omega)\\[1mm]
\bigl(A_n^*\phi_p^{(n)*}(z)-B_n^*\varphi_p^{(n)}(z)\bigr)\chi^{(+)}_{p\ell m}(\Omega)
\end{pmatrix}, & z=z_L.
\end{cases}
\end{equation}

\begin{equation}
\label{mode_NR_down}
\Psi^{-(R)}_{\down,n}(z,\Omega)=
\begin{cases}
\dfrac{1}{N_n}
\begin{pmatrix}
-\varphi_p^{(n)*}(z)\,\chi^{(-)}_{p\ell m}(\Omega)\\[1mm]
\phi_p^{(n)*}(z)\,\chi^{(-)}_{p\ell m}(\Omega)
\end{pmatrix}, & z=z_R,\\[5mm]
\dfrac{1}{N_n}
\begin{pmatrix}
\bigl(A_n^*\phi_p^{(n)*}(z)-B_n^*\varphi_p^{(n)}(z)\bigr)\chi^{(-)}_{p\ell m}(\Omega)\\[1mm]
\bigl(A_n^*\varphi_p^{(n)*}(z)+B_n^*\phi_p^{(n)}(z)\bigr)\chi^{(-)}_{p\ell m}(\Omega)
\end{pmatrix}, & z=z_L.
\end{cases}
\end{equation}

\begin{equation}
\label{mode_NL_up}
\Psi^{-(L)}_{\up,n}(z,\Omega)=
\begin{cases}
\dfrac{1}{N_n}
\begin{pmatrix}
\bigl(-A_n^*\phi_p^{(n)*}(z)+B_n^*\varphi_p^{(n)}(z)\bigr)\chi^{(+)}_{p\ell m}(\Omega)\\[1mm]
\bigl(A_n^*\varphi_p^{(n)*}(z)+B_n^*\phi_p^{(n)}(z)\bigr)\chi^{(+)}_{p\ell m}(\Omega)
\end{pmatrix}, & z=z_R,\\[5mm]
\dfrac{1}{N_n}
\begin{pmatrix}
\varphi_p^{(n)*}(z)\,\chi^{(+)}_{p\ell m}(\Omega)\\[1mm]
\phi_p^{(n)*}(z)\,\chi^{(+)}_{p\ell m}(\Omega)
\end{pmatrix}, & z=z_L.
\end{cases}
\end{equation}

\begin{equation}
\label{mode_NL_down}
\Psi^{-(L)}_{\down,n}(z,\Omega)=
\begin{cases}
\dfrac{1}{N_n}
\begin{pmatrix}
\bigl(A_n^*\varphi_p^{(n)*}(z)+B_n^*\phi_p^{(n)}(z)\bigr)\chi^{(-)}_{p\ell m}(\Omega)\\[1mm]
\bigl(-A_n^*\phi_p^{(n)*}(z)+B_n^*\varphi_p^{(n)}(z)\bigr)\chi^{(-)}_{p\ell m}(\Omega)
\end{pmatrix}, & z=z_R,\\[5mm]
\dfrac{1}{N_n}
\begin{pmatrix}
\phi_p^{(n)*}(z)\,\chi^{(-)}_{p\ell m}(\Omega)\\[1mm]
\varphi_p^{(n)*}(z)\,\chi^{(-)}_{p\ell m}(\Omega)
\end{pmatrix}, & z=z_L ,
\end{cases}
\end{equation}
where the analytic-continuation coefficients are
\begin{align}
A_{n}&=\frac{\sinh(\pi \mu_{n})}{\cosh(\pi p)},
\label{A_def}\\[1mm]
B_{n}&=e^{-\pi p}\,
\frac{\Gamma\!\left(\frac12-ip\right)^2} {\Gamma\!\left(\frac12-ip-i\mu_{n}\right)\Gamma\!\left(\frac12-ip+i\mu_{n}\right)} .
\label{B_def}
\end{align}

Although  \cite{Kanno2017-zv} adopts the opposite metric signature, the redefinition $\gamma^\mu_{(+---)}=-i\,\gamma^\mu_{(-+++)}$ implies $i\slashed{\nabla}_{(+---)}=\slashed{\nabla}_{(-+++)}$ for the full covariant Dirac operator. The corresponding change of Clifford-algebra convention is accompanied by the associated spin-connection convention, so that the first-order Dirac equation and the radial system retain the same functional form as in \cite{Kanno2017-zv}, up to convention-dependent phases and the replacement $m\to M_n$.

\section{Nonlocal Bunch--Davies vacuum and observables}\label{sec4}
\subsection{Bunch--Davies construction in the nonlocal theory}
The local construction of \cite{Kanno2017-zv} determines the BD vacuum through the Euclidean analyticity of the spinor modes and the Bogoliubov transformation between the Euclidean and $R/L$ open chart bases. In the nonlocal theory, this structure is preserved, while its mass-dependent quantities are modified by the branch-dependent replacement $m\to M_n$.

The positive-frequency modes are still selected by analyticity on the lower hemisphere of the Euclidean continuation. Since the nonlocal
equation is solved by local Dirac modes with effective mass $M_n$, the Euclidean analyticity argument carries over directly to real branches, whereas complex branches are understood as formal analytic-continuations. Thus, nonlocality preserves the geometric definition of the BD state while modifying the mass-dependent quantities entering its construction.

In particular, the replacement $m\to M_n$, with $M_n$ determined by Eq.~(\ref{mass_condition}), modifies the hypergeometric indices in Eqs.~(\ref{phi_def})--(\ref{varphi_star_def}), the normalization factor $N_n$ in Eq.~(\ref{Nb_def}), and the analytic-continuation coefficients $A_n$ and $B_n$ in Eqs.~(\ref{A_def}) and (\ref{B_def}). For a real positive-residue branch, the positive- and negative-frequency modes are related by
\begin{equation}
\varphi \to A_n\phi+B_n\varphi^*, \qquad \phi\to -A_n\varphi+B_n\phi^*,
\end{equation}
with all coefficients evaluated at the effective mass $M_n$. Consequently, the Euclidean matching between the $R$ and $L$ charts becomes branch-dependent, modifying the late-time behavior, relative phases of the spinor components, and mixing between the two charts.

The Bogoliubov transformation relating the BD and open chart vacua is modified accordingly. For each real positive-residue branch, its structure remains that of the local theory~\cite{Kanno2017-zv}, with $A\to A_n$ and $B\to B_n$:
\begin{equation}
\label{mij_nonlocal}
m^{(n)}_{ij}=-\frac{B_n^*}{1+A_n^2} \begin{pmatrix} A_n & -1\\ 1 & A_n
\end{pmatrix}_{ij}.
\end{equation}

Defining
\begin{equation}
\omega_n\equiv m_{RR}^{(n)} =m_{LL}^{(n)} =-\frac{A_nB_n^*}{1+A_n^2}, \qquad \zeta_n \equiv m_{RL}^{(n)} =-m_{LR}^{(n)} =\frac{B_n^*}{1+A_n^2},
\end{equation}
the diagonalizing Bogoliubov parameter is
\begin{equation}
\label{gamma_nonlocal}\gamma_{p,n}=\frac{1}{2\zeta_n} \left[ \omega_n^2+\zeta_n^2-1 + \sqrt{ \left(\omega_n^2+\zeta_n^2-1\right)^2 +4\zeta_n^2 } \right],
\end{equation}
where the physical root is chosen by continuity with the local BD solution and by the corresponding canonical Bogoliubov prescription.

\subsection{Branchwise Bunch--Davies vacuum and observables}
For each real positive-residue branch, the BD vacuum takes the same form as in the local theory but with the branch-dependent parameter $\gamma_{p,n}$,
\begin{equation}
\label{BD_branch_state}
|0\rangle^{(n)}_{\mathrm{BD}} = N_{\gamma_{p,n}}^{-1} \exp\!\left( \gamma_{p,n}\,\tilde c_{R,n}^\dagger \tilde d_{L,n}^\dagger +\gamma_{p,n}\,\tilde d_{R,n}^\dagger \tilde c_{L,n}^\dagger \right) |0\rangle^{(n)}_{R'}|0\rangle^{(n)}_{L'},
\end{equation}
where $\tilde c_{R/L,n}$ and $\tilde d_{R/L,n}$ denote the fermionic annihilation operators in the diagonal $R'$- and $L'$-chart basis obtained after the Bogoliubov rotation, $N_{\gamma_{p,n}}=1+|\gamma_{p,n}|^2$, and the reduced density matrix in one open chart inherits the local form,
\begin{equation} \label{rho_nonlocal_branch}
\rho_{R}^{(n)} = \frac{1}{\left(1+|\gamma_{p,n}|^2\right)^2} \left(|0_c\rangle\langle 0_c|+|\gamma_{p,n}|^2|1_c\rangle\langle 1_c|\right)
\otimes \left(|0_d\rangle\langle 0_d|+|\gamma_{p,n}|^2|1_d\rangle\langle 1_d|\right),
\end{equation}
so the mean occupation number for each fermionic sector becomes
\begin{equation} \label{number_branch}
\langle N_{p,n}\rangle= \frac{|\gamma_{p,n}|^{2}}{1+|\gamma_{p,n}|^{2}}.
\end{equation}

Thus, the nonlocal modification of the particle content and entanglement is encoded in $\gamma_{p,n}$. In the occupation-number basis $\left\{|0_c0_d\rangle, |1_c0_d\rangle, |0_c1_d\rangle, |1_c1_d\rangle \right\}$, the density matrix is diagonal, with eigenvalues
\begin{equation}
\lambda_1 = \frac{1}{(1+|\gamma_{p,n}|^2)^2}, \quad \lambda_2 = \lambda_3 = \frac{|\gamma_{p,n}|^2}{(1+|\gamma_{p,n}|^2)^2}, \quad \lambda_4 = \frac{|\gamma_{p,n}|^4}{(1+|\gamma_{p,n}|^2)^2}.
\end{equation}

Indeed, $\Sigma\lambda_i= 1$, as required. The entanglement entropy between the two open charts is defined by the von Neumann entropy of the reduced density matrix. Using the eigenvalues above,
\begin{align}
S_{p,n} =-\sum_{i=1}^{4}\lambda_i\ln\lambda_i = 2 \left[ \ln\left(1+|\gamma_{p,n}|^2\right) - \frac{|\gamma_{p,n}|^2} {1+|\gamma_{p,n}|^2} \ln |\gamma_{p,n}|^2 \right].
\label{entropy_gamma}
\end{align}
Equivalently, in terms of the mean occupation number in Eq.~\eqref{number_branch},
\begin{equation}
S_{p,n} = -2 \left[\left(1-\langle N_{p,n}\rangle\right) \ln\left(1-\langle N_{p,n}\rangle\right) + \langle N_{p,n}\rangle \ln\langle N_{p,n}\rangle \right].
\label{entropy_N}
\end{equation}

The factor of two accounts for the independent particle and antiparticle sectors. Accordingly, $S_{p,n}\to0$ as $|\gamma_{p,n}|\to0$, whereas $|\gamma_{p,n}|^2=1$ yields the maximal entropy $S_{p,n}=2\ln2$.

\subsection{Pole structure and physical branch selection}\label{sec:physical_branch}

For the exponential form factors introduced in Sec.~\ref{sec2}, the effective masses $M_n$ are expressed in terms of branches of the Lambert-$W$ function~\cite{Corless:1996zz}. We now analyze their pole positions and residues~\cite{Buoninfante:2018mre} to identify the physical fermionic sector.

For Model I, defined by Eq.~\eqref{f1}, the characteristic equation is $M e^{iM/\Lambda}=m$, with solutions $M_n=-i\Lambda W_n(im/\Lambda)$. The principal branch $n=0$ is continuously connected to the local Dirac pole as $\Lambda\to\infty$, but is generically complex at finite $\Lambda$. The nonprincipal branches are also complex and are therefore regarded as part of the analytic structure of the propagator rather than as modes admitting a standard particle interpretation.

For Model II, defined by Eq.~\eqref{f2}, the characteristic equation is $M e^{-M/\Lambda}=m$, with solutions $M_n=-\Lambda W_n(-m/\Lambda)$. For $0<m/\Lambda<1/e$, the branches $n=0$ and $n=-1$ are real. The residue at a simple pole is proportional to $\mathcal{R}_n\propto e^{M_n/\Lambda}(1-M_n/\Lambda)^{-1}$. For the principal branch, $0<M_0<\Lambda$ and therefore $\mathcal{R}_0>0$. By contrast, $M_{-1}>\Lambda$ gives $\mathcal{R}_{-1}<0$, corresponding to a ghostlike excitation. At $m/\Lambda=1/e$, the two real branches coalesce at $M=\Lambda$, where the simple-pole residue analysis ceases to apply. All remaining Lambert-$W$ branches are complex. The results are summarized in Table~\ref{tab:poles_residues}.
\begin{table}[H]
\caption{Pole structure and residue classification of the two nonlocal fermionic models.}
\label{tab:poles_residues}
\begin{ruledtabular}
\begin{tabular}{cccc}
Model & Branch & Pole/residue character & Interpretation \\
\hline
I & $n=0$ & complex pole & Local-continuation pole \\
I & $n\neq0$ & complex poles & Nonprincipal complex poles \\[2pt]
\hline
II & $n=0$ & real, positive residue & Physical fermionic pole \\
II & $n=-1$ & real, negative residue & Ghostlike excitation \\
II & $n\neq0,-1$ & complex poles & Additional complex poles
\end{tabular}
\end{ruledtabular}
\end{table}
This classification is also consistent with the Hermiticity condition \eqref{6}. Model II satisfies this condition and admits a real, positive-residue principal branch, whereas Model I is non-Hermitian and has generically complex poles at finite $\Lambda$. 

\subsubsection{Physical Bunch--Davies vacuum and supercurvature}

We therefore restrict the physical analysis to the principal branch $n=0$ of Model II, defined by Eq.~(\ref{f2}). All nonlocal modifications of the BD observables are then encoded in $m\longrightarrow M_0=-\Lambda W_0(-m/\Lambda)$, with $0<m/\Lambda<1/e$.

Accordingly, the physical nonlocal field admits the standard mode expansion
\begin{equation}
\label{field_expansion}
\Psi_{\mathrm{phys}}(x) = \int_0^\infty dp \sum_{\ell,m} \sum_{s=\uparrow,\downarrow} \left[a_{ps\ell m}\, \Psi^{+(\mathrm{BD})}_{ps\ell m}(x;M_0) + b^\dagger_{ps\ell m}\, \Psi^{-(\mathrm{BD})}_{ps\ell m}(x;M_0) \right].
\end{equation}
No independent creation and annihilation operators are assigned to the negative-residue or complex branches.

The physical nonlocal BD vacuum is therefore
\begin{equation}
|0\rangle^{\mathrm{phys}}_{\mathrm{BD,NL}} \equiv |0\rangle^{(0)}_{\mathrm{BD}},
\label{physical_BD_vacuum}
\end{equation}
where $|0\rangle^{(0)}_{\mathrm{BD}}$ denotes the branchwise vacuum of Sec.~\ref{sec4} evaluated at $M_n=M_0$. No tensor-product Hilbert space over the different Lambert-$W$ branches is assumed.

The absence of physical fermionic supercurvature modes found in the local theory~\cite{Kanno2017-zv} is preserved. Setting $p=i\widetilde p$, with $\widetilde p>0$, the analytic-continuation coefficient Eq.~(\ref{B_def}) becomes
\begin{equation}
B_0(i\widetilde p)=e^{-i\pi\widetilde p}\frac{ \Gamma\!\left(\frac12+\widetilde p\right)^2 }{ \left| \Gamma\!\left(\frac12+\widetilde p+i\mu_0\right) \right|^2} \neq0,
\label{B0_supercurvature}
\end{equation}
where $\mu_0=M_0/H$. Hence, the divergent contribution at the opposite boundary cannot be eliminated, and the corresponding mode is not normalizable. Thus, no physical fermionic supercurvature modes arise on the principal branch of Model II.

\subsubsection{Nonlocal corrections to Bunch--Davies observables} \label{sec:NL_BD_corrections}

We now specialize the general results of Sec.~\ref{sec4} to the physical principal branch. In the large-$\Lambda$ regime,
\begin{equation}
\delta M\equiv M_0-m=\frac{m^2}{\Lambda} +\frac{3m^3}{2\Lambda^2} +\mathcal O(\Lambda^{-3}).
\label{M0_expansion}
\end{equation}
The analytic-continuation coefficients of Sec.~\ref{sec3} then become
\begin{equation}
A_0=A(m)+\delta M\,A'(m)+\mathcal O(\delta M^2), \qquad B_0 = B(m)+\delta M\,B'(m)+\mathcal O(\delta M^2).
\end{equation}
Consequently, defining $q\equiv|\gamma_{p,0}|^2$, and $q_p\equiv|\gamma_p(m)|^2$, we obtain
\begin{equation}
q=q_p +\delta M\frac{\partial q_p}{\partial m} +\mathcal O(\delta M^2),
\label{q_expansion}
\end{equation}
where the derivatives with respect to $m$ are evaluated at fixed $p$ and $H$. Using Eqs.~\eqref{number_branch} and \eqref{M0_expansion}, we define the nonlocal correction to the mean occupation number as $\Delta N_p \equiv \langle N_p\rangle_{\mathrm{NL}} - \langle N_p\rangle_{\mathrm{local}}$. To leading order,
\begin{equation}
\Delta N_p= \frac{m^2}{\Lambda(1+q_p)^2} \frac{\partial q_p}{\partial m} +\mathcal O(\Lambda^{-2}). 
\label{DeltaN}
\end{equation}

Similarly, using Eq.~\eqref{entropy_gamma}, we define $\Delta S_p\equiv S_p^{\mathrm{NL}}-S_p^{\mathrm{local}}$. Thus, $\Delta S_p>0$ ($\Delta S_p<0$) corresponds to an enhancement (suppression) of the entanglement between the $R$ and $L$ open charts. Expanding about the local result gives
\begin{align}
\Delta S_p &= \left.\frac{dS}{dq}\right|_{q=q_p}\!\!\!\!\!\! \delta M\,\frac{\partial q_p}{\partial m} + \mathcal O(\delta M^2) = -\frac{2\ln q_p}{(1+q_p)^2} \frac{m^2}{\Lambda} \frac{\partial q_p}{\partial m} +\mathcal O(\Lambda^{-2}).
\label{Delta_entropy}
\end{align}

Since $\delta M>0$ in the large-$\Lambda$ regime, the leading corrections to both the occupation number and the entanglement entropy are controlled by the mass dependence of the local squeezing parameter. Both effects therefore originate from the same nonlocal mass shift $m\to M_0$.
\begin{figure}[H]
\centering
\includegraphics[width=0.5\textwidth]{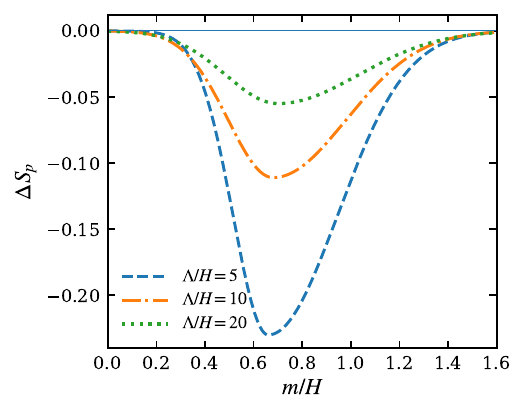}
\vspace*{-0.7cm}
\caption{Leading nonlocal correction $\Delta S_p$ as a function of $m/H$ for $p=0.1$ and different values of $\Lambda/H$, evaluated on the physical principal branch ($n=0$) of Model II.}
\label{fig:DeltaS}
\end{figure}
Figure~\ref{fig:DeltaS} shows that $\Delta S_p<0$ for the parameters considered, indicating a suppression of the $R$--$L$ entanglement. The suppression increases as $\Lambda/H$ decreases and vanishes in the local limit $\Lambda/H\to\infty$.

\section{Conclusion}\label{sec5}
We formulated a nonlocal spin-$1/2$ field theory in curved spacetime with entire function of the covariant Dirac operator. Two exponential form factors were considered, leading to distinct analytic and spectral structures. We also introduced a ${\rm U}(1)$ gauge interaction and showed that the nonlocal corrections can be systematically organized through the Zassenhaus expansion.

As an application, we studied the theory in the open chart of de Sitter spacetime, where the nonlocal Dirac equation can be mapped, branch by branch, onto the local equation with an effective mass $M_n$ determined by the Lambert-$W$ function. The analysis of the propagator poles and their residues identifies the principal branch of Model II as the physical sector for $0<m/\Lambda<1/e$, while the second real branch is ghostlike and the remaining branches are complex.

In the physical sector, nonlocality preserves the Euclidean analyticity of the BD state while modifying the Bogoliubov coefficients, squeezing parameter, particle occupation number, and entanglement entropy through $m\to M_0$. No physical fermionic supercurvature modes arise, and, for the parameters considered, nonlocality suppresses the $R$--$L$ entanglement. The local Dirac theory is smoothly recovered in the limit $\Lambda\to\infty$.

As a continuation of this work, we intend to address other topics in forthcoming studies, such as nonlocal effects in the near-horizon geometry of black holes and the solution of the nonlocal Dirac equation in braneworld scenarios.

\end{document}